\documentclass[12pt]{iopart}
\usepackage{iopams} 
\usepackage{braket}
\usepackage{color}
\usepackage{graphicx}
\usepackage{color,soul}

\newcommand{\oi}{\ensuremath{\omega_\mathrm{i}}}
\newcommand{\os}{\ensuremath{\omega_\mathrm{s}}}

\begin{document}

\title[Probing spectral-temporal correlations with an integrated source of PDC states]{Probing spectral-temporal correlations with a versatile integrated source of parametric down-conversion states} 

\author{Vahid Ansari, Benjamin Brecht, Georg Harder, and Christine Silberhorn}
\address{Integrated Quantum Optics, University of Paderborn, Germany}
\ead{vahid.ansari@uni-paderborn.de}
\vspace{10pt}
\begin{indented}
\item[]\today
\end{indented}

\begin{abstract}
The spectral-temporal correlation and the correlation time of a biphoton wavepacket generated in the process of parametric down-conversion (PDC), is of great importance for a broad range of quantum experiments.
We utilise an integrated PDC source to generate biphotons with different types of spectral-temporal correlations and probe their respective correlation times. The outcomes confirms that the correlation time is independent of the coherence time of the pump light, and it is only determined by the waveguide length and its dispersion properties. Furthermore, we investigate the properties of the PDC biphoton wavepacket exhibiting different types of spectral-temporal correlations and their suitability for quantum-enhanced applications.
\end{abstract}

%
%
%
%
%

\section{Introduction}

Parametric down-conversion (PDC) \cite{Akhmanov:1967vb, Magde:1967bw} is a well-established experimental tool for the generation of photon pair states, which are deployed in various experiments ranging from fundamental tests of quantum mechanics \cite{Ou:1988gk, Weihs:1998cc, Bouwmeester:1999jq} to elaborate quantum information processing protocols \cite{Mattle:1996db, Bouwmeester:1997ji, Jennewein:2000ke, Cuevas:2013gs}. 
In this context, interesting traits of PDC states are, on the one hand, the spectral-temporal correlation between the generated photons and, on the other hand, the biphotons \textit{correlation time} $T_\mathrm{c}$. In the past few years, considerable effort has been focused to control and manipulate the PDC biphoton correlations to realise specifically tailored states exploiting different methods, e.g. in bulk PDC sources by manipulating attributes of the pump field such as beam waist \cite{mosley:2008ki}, spatial chirp and profile \cite{TorresCompany:2009eb} and coherence time \cite{Kuzucu:2005hd} in bulk crystal PDC sources, or the use of different crystals \cite{Shimizu:2009cr, Fedorov:2007cm} and tailored phasematching \cite{Nasr:2008gu, BenDixon:2013jk}.

The correlation time of PDC biphotons is typically in the range of a few hundreds of femtoseconds and measures not only the degree of simultaneity between the pair photons, but also defines the shortest timing information extractable from the PDC state. Thus, the correlation time poses an inherent limitation to the achievable precision of quantum-enhanced applications such as quantum clock synchronisation \cite{Giovannetti:2001dg, Giovannetti:2001eb}, quantum optical coherence tomography \cite{Abouraddy:2002hm} or quantum interferometric optical lithography \cite{Boto:2000eg}. A thorough investigation of its underlying physics is indispensable for pushing those applications further towards their ultimate limits.

With state-of-the-art single photon detectors it is not possible to directly measure $T_\mathrm{c}$, due to their limited timing resolution of at least several tens of picoseconds \cite{Friberg:1985dc}. However, Hong, Ou and Mandel (HOM) showed in their seminal paper that this obstacle can be overcome, when deploying quantum interference of photons at a beamsplitter \cite{Hong:1987gm}. With this method, they measured a correlation time $T_\mathrm{c}$ of about 100 fs. More precisely, they sent the two generated PDC photons into the input ports of a balanced beamsplitter and measured the coincidence events between its output ports when delaying one of the photons with respect to the other. When two photons are identical in all degrees of freedom (arrival time, spectrum and polarisation), they bunch together and emerge from the same output port. Consequently the detected coincidence rate drops to zero. Making the photons distinguishable, for instance by introducing a relative time delay, degrades the quantum interference and the typical HOM dip can be observed. The amount of delay, which is required to render the photons completely distinguishable is proportional to the correlation time $T_\mathrm{c}$. Typically, due to experimental considerations, a narrow-band spectral filtering of PDC photons is applied prior to the detection \cite{Hong:1987gm, Rubin:1994kb}. However, the use of narrow-band filters will alter the created PDC state and thus the width of interference pattern only samples the bandwidth of the applied filters and not the true PDC correlation time.

In this paper we utilise a waveguided PDC source developed in our group which is capable of generating PDC biphotons with well-controlled, tunable spectral-temporal correlations \cite{Eckstein:2011gp, Harder:2013hk}. We probe their correlation times $T_\mathrm{c}$ and show that the properties of the PDC pump field, including its coherence time and chirp, have actually no impact on the correlation time $T_\mathrm{c}$ of the generated biphotons. Then, we present a special case where $T_\mathrm{c}$ is in fact larger than the coherence time of the pump. We give an intuitive explanation for this by analytical calculations which are valid for a wide range of PDC sources and can readily be extended to any PDC. Our results allow us to identify the potential \textit{gain} or \textit{loss} of timing information relative to the pump field for different PDC sources.

\section{Theoretical background}

Earlier theoretical studies have shown that the correlation time of the PDC biphotons does not depend on the spectral bandwidth of the pump \cite{Hong:1985cw, URen:2007fe}. 
Here, we extend these studies by considering chirped pump pulses, which is realistic with consideration of the experimental conditions. We derive the interference pattern between PDC photons and write $T_\mathrm{c}$ only in terms of crystal properties.
Following the standard quantum first order perturbative approach we write the PDC state as \cite{Grice:1997ht, URen:2005wb}
\begin{equation}
  \ket{\psi}_\mathrm{PDC} \propto \ \int \int d\os d\oi \; f(\os,\oi) \; \hat a_{\mathrm s}^\dagger(\os) \hat a_{\mathrm i}^\dagger(\oi) \ket{0},
  \label{eq:state1}
\end{equation}
where the $\hat{a}_\mathrm{s}^\dagger(\os)$ and $\hat{a}_\mathrm{i}^\dagger(\oi)$ are the standard creation operators for a signal photon at frequency $\os$ and an idler photon at frequency $\oi$, respectively. The joint spectral amplitude (JSA) function $f(\os,\oi)$ is a complex-valued function that describes the spectral-temporal structure of the created photon pair. The JSA function comprises the pump envelope function $\alpha(\os+\oi)$ and the phasematching function $\mathrm\phi(\os,\oi)$, which reflect energy and momentum conservation of the PDC process, respectively.
For our analysis, we assume a Gaussian pump spectrum and allow for a possible frequency chirp which is typically present in experimental settings due to the dispersion of optical components. Hence we write
\begin{equation}
  \alpha(\nu_s+\nu_i)=e^{-\left(\frac{\nu_{\mathrm s}+\nu_{\mathrm i}}{\sigma_{\mathrm p}}\right)^2} e^{\imath \beta \left(\nu_{\mathrm s}+\nu_{\mathrm i}\right)^2},
  \label{eq:pump1}
\end{equation}
where $\sigma_{\mathrm p}$ is the spectral width of the pump pulse and $\beta$ characterizes its chirp at the input of the nonlinear medium. Moreover we defined the frequency detunings $\nu_\mu=\omega_\mu-\omega_\mu^{0}$ (with $\mu \in \{ \mathrm{s}, \mathrm{i}\}$) from the perfectly phasematched central PDC frequency $\omega^0_\mathrm{\mu}$. 

The phasematching function is governed by the dispersion properties and the length of the nonlinear waveguide $L$, and is given by
\begin{equation}
  \mathrm\phi(\os,\oi) = e^{-\gamma \left(\frac{L}{2} \Delta k(\os,\oi) \right)^2} e^{\imath \frac{L}{2} \Delta k(\os,\oi)},
  \label{eq:pump2}
\end{equation}
where $\Delta k(\os,\oi)$ is the phase-mismatch between pump, signal and idler in the waveguide. 
Here the sinc-profile of the phasematching function is approximated with a Gaussian function. 
This can be realised by appropriate engineering of the nonlinearity of the waveguide \cite{Branczyk:2011kh, BenDixon:2013jk}.
Using a Taylor expansion up to the first order around the perfectly phasematched frequencies $\omega^{0}_{\mathrm{\mu}}$, we rewrite the phase-mismatch as $L\Delta k\approx \tau_{\mathrm s}\nu_{\mathrm s} + \tau_{\mathrm i}\nu_{\mathrm i}$. Here $\tau_\mu = L(u_\mu^{-1}-u_{\mathrm p}^{-1})$, where $u$ denotes the group velocity of the corresponding fields \cite{URen:2005wb}. In our calculations we dismiss linear phase terms in $\nu_\mu$ in the JSA function, since they only cause a temporal shift of the biphoton state.
Following these calculation we can express the JSA as \cite{SanchezLozano:2011iu, JeronimoMoreno:2009bn}

\begin{equation}
  f(\nu_{\mathrm s},\nu_{\mathrm i})=\exp{ \left[-\left( T^2_{\mathrm{ss}}\nu^2_s + T^2_{\mathrm{ii}}\nu^2_{\mathrm i} + 2 T^2_{\mathrm{si}} \nu_{\mathrm s}\nu_{\mathrm i}\right)\right] },
  \label{eq:jsa_a}
\end{equation}
where we defined

\begin{equation}
  T_{\lambda\mu}^2\equiv\frac{1}{\sigma^2}+\frac{\gamma}{4} \tau_\lambda \tau_\mu - \imath \beta  \qquad \lambda,\mu=\mathrm s, \mathrm i.
  \label{eq:jsa_b}
\end{equation}

To derive the HOM interference pattern, we take the PDC biphoton state in Eq. (\ref{eq:state1}) and apply a time delay $\tau$ to one input arm of the beamsplitter. The rate of coincidence events between the output ports of the beamsplitter is then given by \cite{Grice:1997ht}

\begin{equation}
  R_\mathrm{c}(\tau) \propto 1 -  \int \int  \mathrm{d}\omega \mathrm{d}\tilde\omega \; \mathrm{e}^{\imath(\omega-\tilde\omega)\tau} f(\omega,\tilde\omega) f^*(\tilde\omega,\omega) .
  \label{eq:p_coinc4}
\end{equation}

For $\tau = 0$, i.e.\ perfect temporal overlap of the photons at the beamsplitter, the integrand of Eq. (\ref{eq:p_coinc4}) measures the symmetry of the JSA function describing the PDC state under the exchange of signal and idler. 
Therefore the visibility of the interference probes the indistinguishability of interfering photons. 
By inserting the JSA function into Eq. (\ref{eq:p_coinc4}) we find

\begin{equation}
  R_\mathrm{c}(\tau) \propto 1 -  \int \int  \mathrm{d}\omega \mathrm{d}\tilde\omega \; \mathrm{e}^{\imath(\omega-\tilde\omega)\tau} \mathrm{e}^{-(A\nu_{\mathrm s}^2+A\nu_{\mathrm i}^2+B\nu_{\mathrm s}\nu_{\mathrm i})} ,
  \label{eq:p_coinc5}
\end{equation}
where we define

\begin{equation}
  A\equiv\frac{2}{\sigma_{\mathrm p}^2}+\frac{\gamma}{4} (\tau_\mathrm{s}^2+\tau_\mathrm{i}^2),\qquad
  B\equiv\frac{2}{\sigma_{\mathrm p}^2}+\frac{\gamma}{2} \tau_\mathrm{s} \tau_\mathrm{i}.
  \label{eq:AB}
\end{equation}

\begin{figure}[b]
\begin{center}
\includegraphics[width=.6\linewidth]{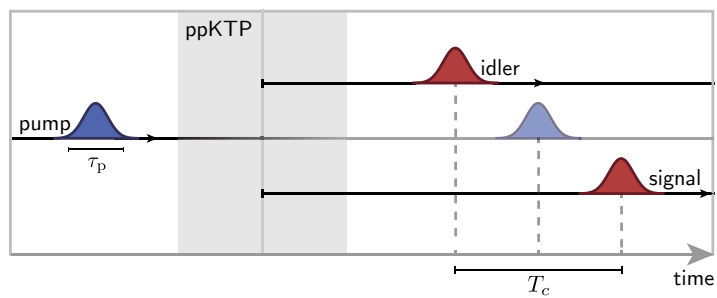}
\caption{\label{fig:gvm} Temporal visualisation of the correlation time of the biphoton. A pump pulse with coherence time $\tau_{\mathrm p}$ decays inside the waveguide and creates daughter photons which travel with different velocities through the medium. On average, the down-conversion process happens at the middle of the waveguide, thus the correlation time $T_\mathrm{c}$ between the down-converted photons is proportional to $L/2$.}
\end{center}
\end{figure}

From Eqs. (\ref{eq:p_coinc5}) and (\ref{eq:AB}) we see that the chirp of the pump field does not affect the interference at all.
Performing the frequency integrations in Eq. (\ref{eq:p_coinc5}) finally yields
\begin{equation}
  R_\mathrm{c}(\tau) \propto 1 - h(\sigma_{\mathrm p},\tau_\mathrm{s},\tau_\mathrm{i}) \
  \mathrm{exp} \left\{ \frac{-\tau^2}{2 \left[ \sqrt{\gamma}\frac{L}{2}(u_\mathrm{s}^{-1}-u_\mathrm{i}^{-1}) \right]^2} \right\} ,
  \label{eq:p_coinc6}
\end{equation}
in correspondence to earlier theoretical studies \cite{Giovannetti:2002hw, URen:2007fe}. Here, $h(\sigma_{\mathrm p},\tau_\mathrm{s},\tau_\mathrm{i})$ is a coefficient which depends on both pump and crystal properties and determines the visibility of the interference. However, the shape of the dip is contained in the exponential function, which is completely independent of the pump. This means that for a given nonlinear crystal, the correlation time $T_\mathrm{c}$ is solely defined by crystal parameters.

We interpret this result in an intuitive manner: in the process of PDC, a pump photon propagates through a waveguide of length $L$ and coherently decays into a pair of photons. These, in turn, travel through the waveguide at different group velocities and thus acquire a temporal walk-off.
Post-selecting on coincidence detection events we find that, on average, the PDC process takes place at the centre of the waveguide. The associated walk-off is the correlation time $T_\mathrm{c}$, which consequently is proportional to half the waveguide length $L/2$, as visualized in Fig. \ref{fig:gvm}. 
This situation simply exemplifies the fact that the correlation time $T_\mathrm{c}$ is independent of the pump coherence time.

\begin{figure}[b]
\begin{center}
\includegraphics[width=.5\linewidth]{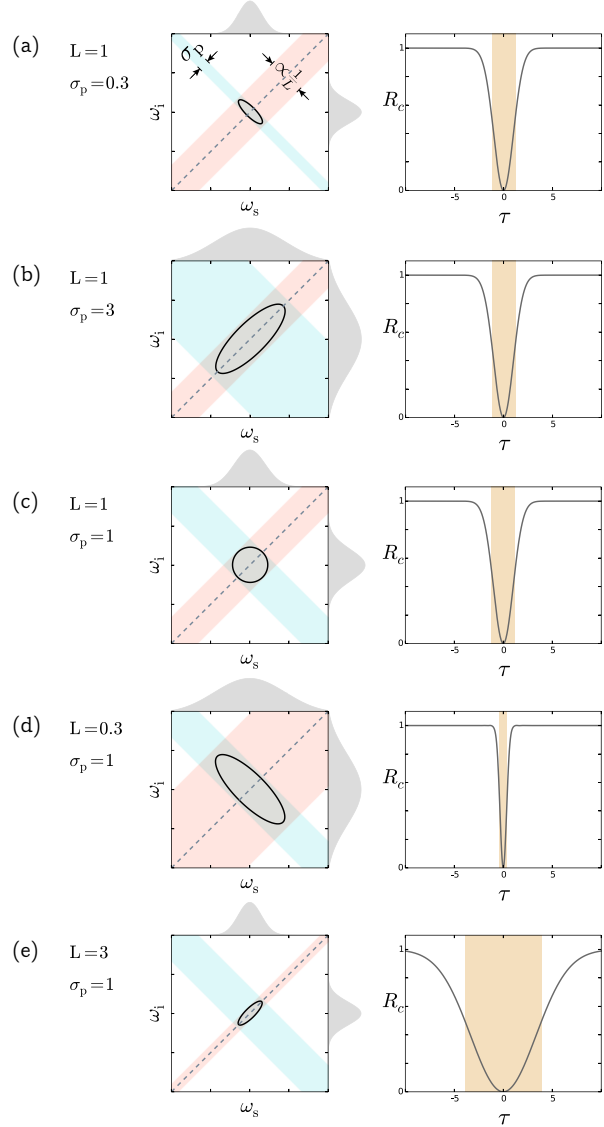}
\caption{\label{fig:jsa_corr} Investigation of the relation between spectral width of the pump field, waveguide length and the correlation time of PDC biphoton wavepacket. All parameters are given in arbitrary units. The JSA reflects PDC processes exhibiting symmetric group velocity matching (SGVM) \cite{URen:2005wb} for different values of the pump spectral width $\sigma_{\mathrm p}$, and the waveguide length $L$ (left). From these we retrieved the HOM dip interference patterns for the respective parameters (right). A detail explanation is presented in the text.}
\end{center}
\end{figure}

We investigate this behaviour in Fig.\ \ref{fig:jsa_corr}, where we plot the JSA function $f(\os,\oi)$ for a waveguide with fixed dispersion properties but at different waveguide lengths $L$ and pump spectral widths $\sigma_{\mathrm p}$ (left) and analyse the resulting HOM interference pattern (right). The grey shaded areas in the JSA plots correspond to signal and idler marginal spectra, whereas the shaded areas in the HOM dip plots highlight the FWHM of the respective interference dip, which by definition is the correlation time $T_\mathrm{c}$. Note that although we focus on a specific waveguide dispersion, our results are generally applicable to any PDC.

In Fig.\ \ref{fig:jsa_corr}-(a,b,c) we assumed a fixed waveguide length $L$ and varied the spectral pump width $\sigma_{\mathrm p}$ and thus the pump coherence time. Despite the different coherence times of the pump field, it is evident that the correlation times of all three biphoton states, as given by the FWHM of the HOM dips, are identical. In contrast, graphs \ref{fig:jsa_corr}-(c,d,e) juxtapose three waveguides with different lengths $L$ that are pumped with similar pump pulses of spectral width $\sigma_{\mathrm p}$. It is obvious that a longer waveguide results in a broader HOM dip, which can be easily understood with the picture of the temporal walk-off between signal and idler.

Furthermore, a comparison between Fig.\ \ref{fig:jsa_corr}-(a,b) shows that in general, the correlation time is not simply the convolution of signal and idler marginal distributions. It is important to recognize that the JSA is fundamentally a two-dimensional distribution function, whereas marginal spectra correspond to the traced-out system. Therefore, the marginals do not contain full information of any biphoton state. Only in the case of a spectrally decorrelated PDC we find that $T_\mathrm{c}$ is proportional to the convolution of the signal and idler marginal distributions \cite{Friberg:1985dc, Hong:1987gm}.

Let us now consider the temporal information of the PDC state. It has been shown that entanglement, including spectral entanglement, can be exploited to enhance the precision of quantum metrology protocols. 
Comparing the cases in Fig. \ref{fig:jsa_corr} (a) and (e), we first note that they have the same amount of spectral entanglement and the same marginal distributions. However, by comparing the resulting HOM interference patterns we realise that spectrally anticorrelated states provide a considerably shorter correlation time. 
To understand this, it is helpful to consider the joint temporal amplitude (JTA) function corresponding to the JSA, which essentially has the shape of the JSA but is rotated about $90^\circ$. The correlation time between signal and idler photons measures the simultaneity of the photons, which denotes $|t_\mathrm{s}-t_\mathrm{i}|$. This corresponds to the width of the JTA function along $-45^\circ$ in the $(t_\mathrm{s},t_\mathrm{i})$-plane. Thus, the spectrally anticorrelated state from Fig. \ref{fig:jsa_corr} (a) is ideal for measuring arrival time differences, whereas the spectrally correlated state from Fig. \ref{fig:jsa_corr} (e) is adapted to measuring absolute times $t_\mathrm{s}+t_\mathrm{i}$ \cite{Giovannetti:2001dg}.

\section{Experimental results and discussion}

\begin{figure}[t]
\begin{center}
\includegraphics[width=.7\linewidth]{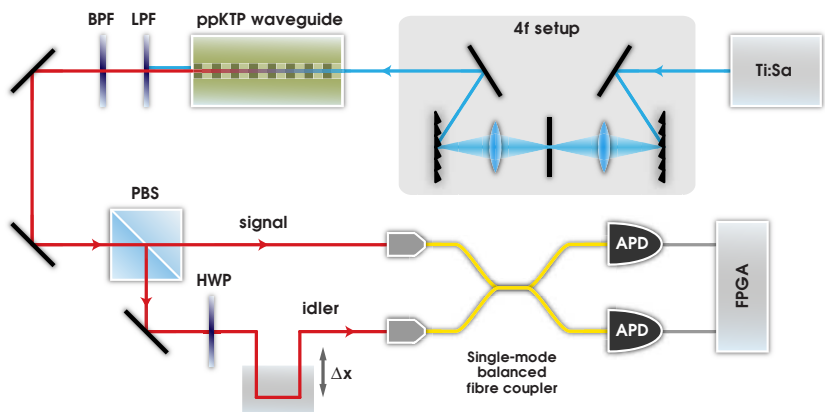}
\caption{\label{fig:setup} Experimental setup. A 4f-setup is employed as tunable spectral filter to control the coherence time of the pump field and thereby the spectral correlation of the biphoton state. An $\mathrm{8 mm}$ long ppKTP waveguide is used as degenerate type-II PDC source. We use a longpass filter (LPF) to block out the pump beam. A broad bandpass filter (BPF) with a transmission band of approximately $\mathrm{8\ nm}$ centred at the central PDC wavelength suppresses the background noise, but does not cut the biphoton spectrum. The PDC photons are separated by a polarising beamsplitter (PBS). A half waveplate (HWP) is used to make the polarisation of signal and idler photons identical. Then the PDC photons are interfered at a balanced fibre coupler.}
\end{center}
\end{figure}

To examine the theory, we use a periodically poled potassium titanyl phosphate (ppKTP) waveguided PDC source recently presented in \cite{Eckstein:2011gp, Harder:2013hk}. 
Our source provides a frequency-degenerate type-II PDC at telecommunication wavelengths, where the phasematching function is oriented along $+59^\circ$ in the $(\os,\oi)$-plane. 
This allows us to realise PDC states with all possible types of spectral correlations, simply by changing the spectral width of the pump pulse, similar to Fig. \ref{fig:jsa_corr}-(a-c).
A schematic of the experimental setup is shown in Fig.\ \ref{fig:setup}. 

We deploy a Ti:Sapphire oscillator with a repetition rate of $\mathrm{80\ MHz}$ which provides ultrafast pulses with duration of $\mathrm{200\ fs}$ at the central wavelength of $\mathrm{767.5\ nm}$. 
To change the spectral width $\sigma_{\mathrm p}$ or equivalently the coherence time of the pump pulses we use a 4f-setup as a tunable spectral filter. The 4f-setup consists of two diffraction gratings, two lenses and one adjustable slit in the centre, as demonstrated in Fig.\ \ref{fig:setup}. This arrangement is a one dimensional Fourier processor. The diffracted light beam is focused on the Fourier plane, where we use a slit to manipulate the spectral width of the pump field. However, this arrangement cuts off the spectrum as a rectangular bandpass filter which distorts the spectrum from its initially Gaussian shape.
Nevertheless, a simple modelling shows that it is possible to recover the Gaussian shape of the spectrum only by tilting the slit. We use this experimental technique to maintain the original shape of the pump pulses.

\begin{table*}[b]
\caption{\label{tab:table1}
The correlation time for spectrally decorrelated and (anti-)correlated PDC states. The measured correlation time $T_\mathrm{c}$, the spectral FWHM of individual fields $\Delta\lambda$, the temporal FWHM of each fields $\Delta\tau$ and the temporal convolution of daughter pulses $\Delta\tau_{\mathrm{conv.}}=\sqrt{ \Delta\tau_{\mathrm{s}}^2 + \Delta\tau_{\mathrm{i}}^2 }$ are listed.}
\begin{tabular}{lcccccc}
\br
Spectral correlation & 
$T_\mathrm{c}\: \mathrm{(ps)}$ & 
$\Delta\lambda_{\mathrm{s/i}}\: \mathrm{(nm)}$ 
& $\Delta\lambda_{\mathrm{p}}\: \mathrm{(nm)}$ 
& $\Delta\tau_{\mathrm{s/i}}\: \mathrm{(ps)}$ 
& $\Delta\tau_{\mathrm{p}}\: \mathrm{(ps)}$ 
& $\Delta\tau_{\mathrm{conv}}\: \mathrm{(ps)}$  \\
\mr
correlated    & $1.21\pm0.03$ & $5.84/10.14$  & $4.5$ & $0.59/0.34$   & $0.19$ & $0.68$ \\
decorrelated  & $1.16\pm0.01$ & $4.3/5.46$  & $2$   & $0.8/0.63$  & $0.43$ & $1.02$ \\
anticorrelated  & $1.10\pm0.02$ & $3.06/3.12$ & $0.7$ & $1.13/1.11$   & $1.24$ & $1.58$ \\
\br
\end{tabular}
\end{table*}

\begin{figure}[t]
\begin{center}
\includegraphics[width=.7\linewidth]{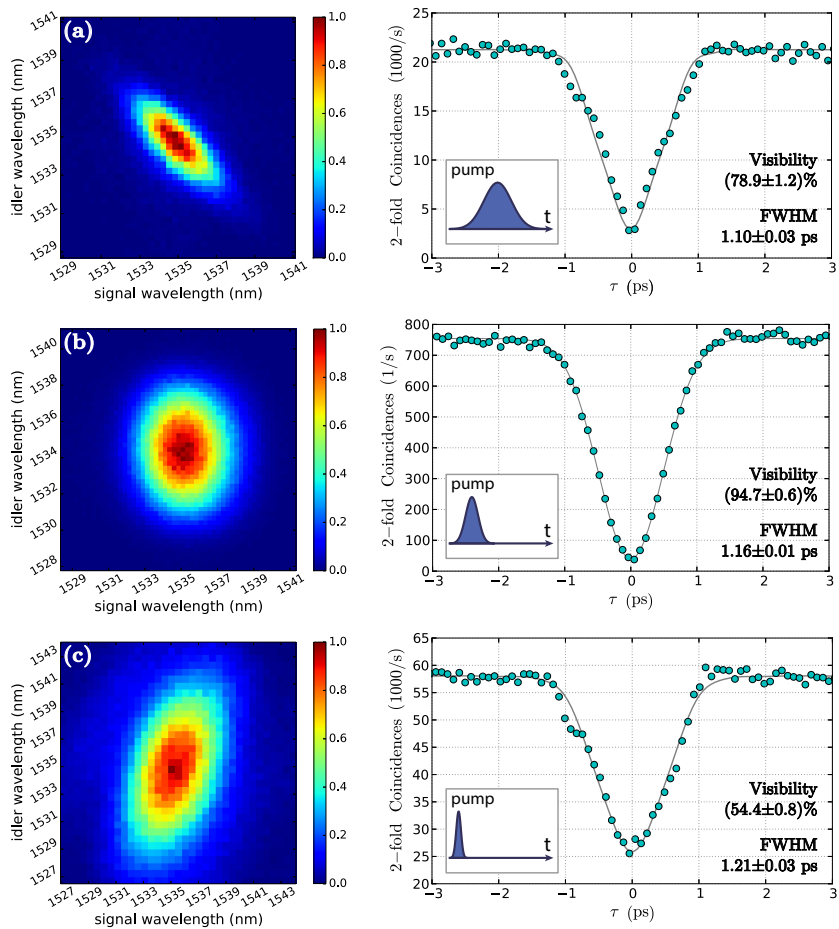}
\caption{\label{fig:jsi} The joint spectral intensity (JSI) and HOM interference measurements for spectrally (a) anticorrelated, (b) decorrelated and (c) correlated PDC states. Left: A fibre spectrometer with the resolution of $\mathrm{1.8\times1.8\ nm^{2}}$ is used to measure the JSI of the PDC biphoton state \cite{Avenhaus:2009hd} for different spectral widths of the pump field $\Delta\lambda_\mathrm{p}$ listed in TABLE \ref{tab:table1}.
Right: The corresponding HOM interference patterns, performed with pump energies as low as 0.6 pJ per pulse leading to a mean photon number of 0.002 pairs per pulse. The error bars are smaller than the points.
In the correlated case we removed the bandpass filter (BPF), since the marginal spectra are broader than the filter transmission in this case.
}
\end{center}
\end{figure}

For the HOM measurement we interfere the signal and idler photons on a single-mode balanced fibre coupler after separating them with a polarizing beamsplitter (PBS) and rotating the signal polarisation with a half-wave plate (HWP). Using a translation stage we introduce a path difference between the interfering photons while recording the coincidence events between the output ports of the coupler with avalanche photo diodes (APDs) connected to a homemade field programmable gate array (FPGA).

In Fig.\ \ref{fig:jsi} we plot the measured joint spectral intensity functions $|f(\os,\oi)|^2$ and the corresponding HOM dips for cases of a spectrally anticorrelated, a decorrelated and correlated PDC \cite{Avenhaus:2009hd}. In agreement with our expectations, the widths of the dips and hence $T_\mathrm{c}$ are nearly identical upon significantly different coherence times of the pump field and are in excellent agreement with the theoretical prediction of $1.16\pm0.03\ $ ps as computed from Eq.\ (\ref{eq:p_coinc6}). 
The small differences of the correlation times can be explained by the inherent sinc-shape of the phasematching function. 
This can be verified by numerical modelling of the exact JSA function and the interference pattern, which is plotted in Fig.\ \ref{fig:jsi} as the grey fit functions.
In TABLE \ref{tab:table1} a summary of the crucial parameters and outcomes of our measurement is presented.

In spectrally decorrelated and correlated states, $T_\mathrm{c}$ is actually larger than the coherence time of the pump $\Delta\tau_\mathrm{p}$.
This corresponds to the schematic in Fig. \ref{fig:gvm}, where the photons are on average detected \textit{outside} the pump. Only for the case of the spectrally anticorrelated PDC, the photons are found \textit{inside} the pump. A similar observation was reported previously in \cite{Kuzucu:2005hd}, where the correlation time of biphoton wavepacket was compared utilising pulse and continuous wave pumping. Moreover we verify that $T_\mathrm{c}$ equals the convolution of the signal and idler marginal durations only for the case of a decorrelated PDC and differs otherwise. 

With these results we can also quantify the \textit{gain} or \textit{loss} of timing information relative to the pump field, by observing the HOM interference between PDC photons. Comparing $T_\mathrm{c}$ to the pump duration $\Delta\tau_\mathrm{p}$, we find that the spectrally anticorrelated PDC state is tailored to precisely retrieve the correlation time $|t_\mathrm{s}-t_\mathrm{i}|$. The most prominent example for this situation is of course the original work of Hong, Ou and Mandel \cite{Hong:1987gm}. On the other hand, the spectrally correlated PDC state is appropriate to measure the sum of arrival times $t_\mathrm{s}+t_\mathrm{i}$ \cite{Giovannetti:2001dg}.

\section{Conclusion}

We theoretically and experimentally investigated the properties of PDC states with different types of spectral-temporal correlations. Using a tunable integrated source, we generated spectrally decorrelated, positively and negatively correlated PDC states. The results shows that the coherence time and chirp of the pump field indeed has no effect on the correlation time of the generated biphoton state. We presented an intuitive model for the origin of the correlation time of PDC photons, which is in complete agreement with our experimental outcomes. Furthermore, the appropriateness of different PDC states in respect to derive timing information were discussed. We shown, depending on the spectral correlation of the PDC state, by measuring the HOM interference one can gain or lose timing information compared with the classical pump field. This fundamental insight can be useful to the diverse number of applications relying on the interference of PDC photons.

\section{Acknowledgements}

We acknowledges helpful discussions with Malte Avenhaus.

\section*{References}
\bibliography{papers}

\end{document}